\documentclass[twocolumn,showpacs,preprintnumbers,amsmath,amssymb,floatfix,nofootinbib]{revtex4}
\usepackage{graphicx,color}
\usepackage{dcolumn}
\usepackage{bm}
\usepackage{amsthm}
\usepackage{amsmath}
\usepackage{amsfonts}

\newcommand{\be}{\begin{equation}}
\newcommand{\ee}{\end{equation}}
\newcommand{\ba}{\begin{eqnarray}}
\newcommand{\ea}{\end{eqnarray}}
\newcommand{\hhh}{\,,\hspace{0.3cm}}
\newcommand{\hnh}{\hspace{0.1cm}}
\newcommand{\non}{\nonumber}
\newcommand{\n}[1]{\label{#1}}
\newcommand{\eq}[1]{(\ref{#1})}

\begin{document}
 
\title{Geometric Properties of Static EMdL Horizons} 

\author{Shohreh Abdolrahimi}
\email{abdolrah@ualberta.ca}
\author{Andrey A. Shoom}
\email{ashoom@ualberta.ca}
\affiliation{Theoretical Physics Institute, University of Alberta, 
Edmonton, AB, Canada,  T6G 2G7}

\begin{abstract}
We study non-degenerate and degenerate (extremal) Killing horizons of arbitrary geometry and topology within the Einstein-Maxwell-dilaton model with a Liouville potential (the EMdL model) in $d$-dimensional ($d\geqslant 4$) static space-times. Using Israel's description of a static space-time, we construct the EMdL equations and the space-time curvature invariants: the Ricci scalar, the square of the Ricci tensor, and the Kretschmann scalar. Assuming that space-time metric functions and the model fields are real analytic functions in the vicinity of a space-time horizon, we study behavior of the space-time metric and the fields near the horizon and derive relations between the space-time curvature invariants calculated on the horizon and geometric invariants of the horizon surface. The derived relations generalize the similar relations known for horizons of static  four and 5-dimensional vacuum and 4-dimensional electrovacuum space-times. Our analysis shows that all the extremal horizon surfaces are Einstein spaces. We present necessary conditions for existence of static extremal horizons within the EMdL model.
 \end{abstract}

\pacs{04.20.-q, 04.50.-h, 04.50.Gh, 04.20.Ex \hfill  
Alberta-Thy-04-11}

\maketitle

\section{INTRODUCTION}
The Einstein-Maxwell-dilaton model with a Liouville potential (the EMdL model) follows from low energy string theories and can be obtained by dimensional reduction of a higher-dimensional Einstein-Maxwell theory with a cosmological constant. Some exact solutions to the EMdL equations, including solutions representing black holes, were derived and analyzed in \cite{7}. Later it was demonstrated that no static, spherically symmetric, asymptotically flat or (anti)-de Sitter ((a)dS) solutions to $d$-dimensional ($d\geqslant 3$) EMdL equations exist \cite{15}. Static, spherically symmetric, non-(a)dS, non-asymptotically flat solutions to $d$-dimensional ($d\geqslant 4$) EMdL equations which represent black holes were discussed in \cite{12}. In particular, three families of black hole solutions were constructed and analyzed. The first family of solutions has two horizons (or one extremal horizon) if $a^2<2/(d-2)$, where $a$ is the dilaton coupling constant (see the action \eq{Action2} below), and there is one horizon if $a^2>2/(d-2)$. A black hole solution does not exist for $a^2=2/(d-2)$. The second family of solutions has only one horizon, while the third family exists only for non-zero value of electric charge. Due to the complicated form of the lapse function an explicit location of the horizons of the third family is not known. However, in 4-dimensional case a single horizon exists for $a=-1$. If the cosmological constant $\Lambda$ is positive and $a^2<1/3$ or $a^2>1$ there can be none, one, two horizons, or an extremal horizon. If $\Lambda<0$ or $1/3<a^2<1$, there is a single horizon. Static topological black hole solutions\footnote[1]{For a review on topological black holes see, e.g. \cite{19,20}. } to 4-dimensional EMdL equations with $\Lambda<0$, whose horizon surface has zero or negative constant curvature, were obtained in \cite{18}, and static black plane solutions were studied in \cite{17}. Static black hole solutions to the $d$-dimensional EMdL model whose horizon surface has zero, positive, or negative constant curvature were constructed in \cite{16}. Black hole solutions to $d$-dimensional EMdL equations were constructed in \cite{1} by using dimensional reduction of $(d+1)$-dimensional black string solutions of the Einstein gravity with a negative cosmological constant and by applying an $SL(2,R)$ transformation to $(d-1)$-dimensional action. These solutions have dilaton field diverging at asymptotic infinity, such that the corresponding Liouville potential vanishes. Cylindrically symmetric solutions to $d$-dimensional EMdL equations were analyzed, and 4-dimensional cylindrically symmetric solutions representing black holes and gravitating solitons were found in \cite{6}. A general $d$-dimensional, cylindrically symmetric solution was obtained for a certain relation between coupling constants. Spherically symmetric, dyonic black hole solutions to $d$-dimensional EMdL equations were derived in \cite{14}.

Having the relatively rich variety of solutions to the EMdL equations which possess regular horizon(s), one is interested to know what types of regular horizons exist within the EMdL model and what are their properties. It is of customary interest to study regular extremal horizons within this model. Regular extremal horizons within other models and their near-horizon geometry were studied, e.g. in \cite{Ha1,Ha1a,Ha2,Ha3}. In particular, in \cite{Ha1} four and $5$-dimensional extremal horizons and their near-horizon geometry were studied within a more general model which includes many abelian vector fields and scalar fields and which can be reduced to the EMdL model. 

In this paper we shall study static Killing horizons of arbitrary geometry and topology within the EMdL model in $d$-dimensional ($d\geqslant 4$) space-times with electrostatic field. By ``static horizons of arbitrary geometry and topology" we mean such horizons whose corresponding space-time can be foliated by $(d-2)$-dimensional equipotential surfaces with no vanishing gradient. 
Given the EMdL action, we are interested to know if there are associated static solutions to the corresponding EMdL equations which have regular Killing horizons. Here we shall consider both non-degenerate and degenerate (extremal) Killing horizons. Such horizons can be horizons associated with black objects or space-time cosmological horizons. To find if such horizons exist, one needs to solve the EMdL equations with no spatial symmetry imposed, which is a formidable problem. Here we shall undertake a less difficult task. Namely, we assume that space-time metric functions and the model fields are real analytic functions in the vicinity of a space-time Killing horizon. Substituting analytic expansions of the metric functions and fields near the horizon into the EMdL equations we derive equations for the expansion coefficients. Solving these equations, we can derive the higher order expansion coefficients in terms of those corresponding to the metric functions and fields defined on the horizon. These expansions allow us to formulate necessary conditions for existence of regular extremal horizons. We call the conditions necessary because the space-time outside a regular extremal horizon may have singularities which are located beyond the radius of convergence of the expansions. Existence or non-existence of such singularities can be established if a global solution is constructed. 

To study geometric properties of regular Killing horizons we restrict ourselves to the space-time curvature invariants: the Ricci scalar, the square of the Ricci tensor, and the Kretschmann scalar. Substituting the expansions into expressions for the space-time curvature invariants we derive relations between the space-time curvature invariants calculated on the Killing horizon and geometric invariants of the horizon surface. One of such relations corresponding to an electrostatic 4-dimensional space-time was derived in \cite{AFS},
\ba\n{A24.IV}
{\cal K}\rvert_{{\cal H}}&=&3\left({\cal R}\rvert_{{\cal H}}+\frac{1}{4}F^2\rvert_{{\cal H}}\right)^2+\frac{1}{8}F^4\rvert_{\cal H}\,.
\ea 
Here ${\cal K}\rvert_{\cal H}$ is the Kretschmann scalar calculated on the space-time horizon ${\cal H}$, ${\cal R}\rvert_{\cal H}$ is the Ricci scalar of the horizon surface, and $F^2\rvert_{\cal H}$ is the electromagnetic field invariant calculated on the horizon. This relation is a generalization of an analogous relation corresponding to a $4$-dimensional vacuum static space-time with a Killing horizon \cite{FS}. It was used in \cite{FS} to calculate vacuum energy density near a static $4$-dimensional black hole using Page's \cite{P} and Brown's \cite{B} approximations. An analogous relation was derived for horizon of a 5-dimensional vacuum static space-time in \cite{ASP}, 
 \ba\n{A24.IVb}
{\cal K}\rvert_{{\cal H}}&=&6{\cal R}_{AB}{\cal R}^{AB}\rvert_{\cal H}\,.
\ea 
Here ${\cal R}_{AB}{\cal R}^{AB}\rvert_{\cal H}$ is the square of the Ricci tensor of the 3-dimensional horizon surface.   
In this paper we derive relations which generalize the relations above to horizons of static $d$-dimensional space-times of the EMdL model. 

Let us note that throughout our paper we shall consider only secondary scalar hair which are induced by the electrostatic field and a Liouville potential, in other words, we shall consider a dilaton field $\varphi$ \cite{Ortin}. 

Our paper is organized as follows: in Sec. II we present the EMdL equations. In Sec. III using Israel's description we define general form of a static space-time metric and bring the EMdL equations to the form which is convenient for our analysis. Section IV contains expressions for the space-time curvature invariants. In Sec. V we analyze space-time geometry near non-degenerate and extremal horizons and derive the relations between the space-time curvature invariants calculated on space-time horizon and geometric quantities of the horizon surface. We shall present necessary conditions for existence of static extremal horizons within the EMdL model. Summary of our results is presented in Sec. VI.

In this paper we use the following convention of units: $G_{(d)}=c=1$, where $G_{(d)}$ is the $d$-dimensional gravitational constant. The space-time signature is $+(d-2)$, and the sign conventions are that adopted in \cite{MTW}.
\section{The EM${\text d}$L equations}
The EMdL model in a $d$-dimensional space-time is defined by the following action:
\ba
&&\hspace{0.5cm}S[g_{\alpha\beta},A_{\alpha},\varphi]=\frac{1}{16\pi}\int d^{d}x \sqrt{-g} ~{\cal L}\,,\non\\
&&\hspace{-0.6cm}{\cal L}=R-\frac{1}{2}g^{\alpha\beta}(\nabla_\alpha\varphi)(\nabla_\beta\varphi)-\frac{1}{4}e^{a\varphi}F^{2}-2\Lambda e^{-b\varphi}\n{Action2}.
\ea
Here $a$, $b$, and $\Lambda$ are arbitrary coupling constants, $\varphi$ is the dilaton field, $F^2= F_{\alpha\beta}F^{\alpha\beta}$, $F_{\alpha\beta}\equiv\nabla_{\alpha}A_{\beta}-\nabla_{\beta}A_{\alpha}$ is the electromagnetic field tensor, where $A_{\alpha}$ is the electromagnetic $d$-potential, and $\Lambda e^{-b\varphi}$ is a Liouville potential which one may consider as an effective ``cosmological constant". Here and in what follows, $\nabla_\alpha$ denotes the covariant derivative defined with respect to a $d$-dimensional metric $g_{\alpha\beta}$.
 
The action \eq{Action2} covers the following cases: the case $b=0$ reduces the Liouville potential to the cosmological constant $\Lambda$; the case $a=0$ is the Einstein-Maxwell theory with the scalar kinetic term and the Liouville potential. In a 10-dimensional space-time, if $F_{\alpha\beta}=0$ the action describes tachyon-free non-supersymmetric string theory (see, e.g., \cite{Tac1,Tac2,Tac3,Tac4,Tac5}). For specific dimension-dependent values of the coupling constants $a$ and $b$ the action \eq{Action2} is the Kaluza-Klein reduction of a $(d+1)$-dimensional general relativity with a cosmological constant and rotation or twist (see, e.g., \cite{RTtheory}).
 
The EMdL equations derived from the action above are the following:
\ba
&&R_{\alpha\beta}-\frac{1}{2}g_{\alpha\beta}R=8\pi T_{\alpha\beta}~,\non\\
\hspace{-0.3cm}8\pi T_{\alpha\beta}&=&\frac{1}{2}\,(\nabla_{\alpha}\varphi)(\nabla_{\beta}\varphi)-\frac{1}{4}\,(\nabla_{\gamma}\varphi)(\nabla^{\gamma}\varphi)g_{\alpha\beta}\non\\
&+&\frac{1}{2}e^{a\varphi}\left(F_{\alpha}^{~\gamma}F_{\beta\gamma}
-\frac{1}{4}F^2g_{\alpha\beta}\right)-e^{-b\varphi}\Lambda g_{\alpha\beta},\n{1.1a.IV}\\
&&\nabla_{\beta}(e^{a\varphi}F^{\alpha\beta})=0\,,\n{1.1c}\\
&&\nabla_{\alpha}\nabla^{\alpha}\varphi=\frac{1}{4}ae^{a\varphi}F^2-2b\Lambda e^{-b\varphi}\n{1.1d}.
\ea
It is convenient to rewrite the Einstein equation \eq{1.1a.IV} in the following form:
\ba
R_{\alpha\beta}&=&\frac{1}{2}(\nabla_{\alpha}\varphi)(\nabla_{\beta}\varphi)+\frac{2\Lambda e^{-b\varphi}}{d-2}g_{\alpha\beta}\non\\
&+&\frac{1}{2}e^{a\varphi}\left(F_{\alpha}^{~\gamma}F_{\beta\gamma}
-\frac{F^2g_{\alpha\beta}}{2(d-2)}\right).\n{1.3b}
\ea
In the next section we present these equations in the form corresponding to a static space-time. 
\section{Static EM$\text{d}$L space-time}
Here we shall focus on static, $d$-dimensional ($d\geqslant4$) space-times of arbitrary geometry and topology (in the sense discussed in the Introduction) and use Israel's description \cite{3,4,5}. Such space-times admit a Killing vector $\xi^{\alpha}=\delta^{\alpha}_{\,\,\, 0}$, ($x^0\equiv t$, and $t$ is coordinate time), which is timelike in the domain of interest: $\xi^{\alpha}\xi_{\alpha}=g_{00}\equiv-k^2<0$, and hypersurface $t=const$ orthogonal. Thus, the space-time metric $g_{\alpha\beta}$, $(\alpha,\beta,...=0,...,d-1)$ can be presented in the following form:
\ba\n{A1.IV}
&&\hspace{-1cm}ds^2=g_{\alpha\beta}dx^{\alpha}dx^\beta=-k^2dt^2+\gamma_{ab}dx^adx^b\,,
\ea
where $\gamma_{ab}=\gamma_{ab}(x^c)$, $(a,b,c,...=1,...,d-1)$, is the metric on a $(d-1)$-dimensional hypersurface $t=const$. 
Assuming that $(\nabla_\alpha k)(\nabla^\alpha k)$ vanishes nowhere in the domain of interest, we can define $k$ as one of the space-time coordinates in the domain and consider $(d-2)$-dimensional equipotential spacelike surfaces $\Sigma_k$ of constant $k$ and $t$. The norm of the spacelike vector $\delta_\alpha^{~k}=\nabla_\alpha k$, which is orthogonal to $\Sigma_k$, is
\be\n{A5.IV}
\kappa^2(k,x^A)\equiv\delta_\alpha^{~k} g^{\alpha\beta} \delta^{~k}_{\beta}=g^{kk}=-\frac{1}{2}(\nabla^{\alpha}\xi^{\beta})(\nabla_{\alpha} \xi_{\beta})\,.
\ee
Thus, the metric \eq{A1.IV} can be written in the following form:
\ba\n{A4.IV}
&&\hspace{-0.5cm}ds^2=-k^2dt^2+\kappa^{-2}dk^2+h_{AB}dx^Adx^B.
\ea
Here $h_{AB}=h_{AB}(k,x^C)$, $(A,B,C,...=2,...,d-1)$ is the metric on $\Sigma_k$. The space-time \eq{A4.IV} has a horizon ${\cal H}$ defined by $k=0$, which is a Killing horizon. The $(d-2)$-dimensional horizon surface is defined by $k=0$ and $t=const$.
In this section we write the EMdL equations \eq{1.1c}-\eq{1.3b} corresponding to the metric \eq{A4.IV}. 

To begin with, we consider $(d-1)$-dimensional hypersurfaces of $t=const$ and use the Gauss and Codazzi relations (see, e.g., \cite{3, Eis})
\ba
\hspace{-0.09cm}R_{a\beta\gamma b}\hnh n^{\beta}n^{\gamma}k&=&\gamma_{ac}{\cal \bar{S}}_{b~,t}^{~c}+\epsilon k_{|ab}+k{\cal \bar{S}}_{ac}{\cal \bar{S}}_b^{~c}\,,\n{CC1a}\\
\hspace{-0.09cm}R_{\alpha bcd}\hnh n^{\alpha}&=&{\cal \bar{S}}_{bc|d}-{\cal \bar{S}}_{bd|c}\,,\n{CC2}\\
\hspace{-0.09cm}R_{abcd}&=&{\bar R}_{abcd}+\epsilon({\cal \bar{S}}_{ad}{\cal \bar{S}}_{bc}-{\cal \bar{S}}_{ac}{\cal \bar{S}}_{bd})\,,\n{CC1}
\ea
where $n^{\alpha}=\xi^{\alpha}/k$ is a unit time-like vector orthogonal to a $(d-1)$-dimensional hypersurface $t=const$, $\epsilon\equiv n^{\alpha}n_{\alpha}=-1$, ${\cal \bar{S}}_{ab}$ is the extrinsic curvature of a hypersurface $t=const$ defined as 
\be
{\cal \bar{S}}_{ab}\equiv\frac{1}{2}k^{-1}\gamma_{ab,t}\,,\n{S4t}
\ee 
and ${\bar R}_{abcd}$ is the Riemann tensor corresponding to the metric $\gamma_{ab}$. Here and in what follows, the barred geometric quantities correspond to $(d-1)$-dimensional hypersurfaces $t=const$, and the stroke stands for the covariant derivative defined with respect to the metric $\gamma_{ab}$.
Contraction of the relations \eq{CC1a}-\eq{CC1} with the use of the metric \eq{A1.IV} gives
\ba
&&\hspace{-0.09cm}R_{\alpha\beta}n^{\alpha}n^{\beta}=-{\cal \bar{S}}_{ab}{\cal \bar{S}}^{ab}-\epsilon k^{-1}k_{|a}^{~|a}-k^{-1}{\cal \bar{S}}_{,t}\,,\n{CC3}\\
&&\hspace{-0.09cm}R_{\alpha b}n^{\alpha}=-{\cal \bar{S}}_{,b}+{\cal \bar{S}}_{b~|c}^{~c}\,,\n{CC4}\\
&&\hspace{-0.09cm}R_{ab}={\bar R}_{ab}-\epsilon {\cal \bar{S}}{\cal \bar{S}}_{ab}-k^{-1}k_{|ab}-\epsilon k^{-1}\gamma_{ac}{\cal \bar{S}}_{b~,t}^{~c}\,,\n{CC5}
\ea
where ${\cal \bar{S}}\equiv\gamma^{ab} {\cal \bar{S}}_{ab}$. 
Since the metric \eq{A1.IV} is static, ${\cal \bar{S}}_{ab}$ vanishes. Thus, the expressions \eq{CC3}-\eq{CC5} imply that for the static space-time \eq{A1.IV} the Ricci tensor components are 
\ba
\hspace{-0.6cm}R_{tt}=k k_{|a}^{~|a}\hhh R_{ta}=0
\hhh R_{ab}={\bar R}_{ab}-k^{-1}k_{|ab}\,.\n{R11}
\ea
Let us now write the components of the $d$-dimensional Riemann and Ricci tensors in terms of geometric quantities corresponding to a $(d-2)$-dimensional surface $\Sigma_k$. Applying the replacements $t\rightarrow k$, $k\rightarrow \kappa^{-1}$, $n^{\alpha}\rightarrow \delta^{\alpha}_{~k} \kappa$, $\epsilon\rightarrow 1$ to the Gauss and Codazzi relations \eq{CC1a}-\eq{CC1} we derive 
\ba
&&\hspace{-0.3cm}R_{AkkB}=\kappa^{-1}\left(h_{AC}{\cal S}_{B\,\,\,,k}^{\,\,\,\,C}+(\kappa^{-1})_{;AB}+\kappa^{-1}{\cal S}_{AC}{\cal S}_{B}^{\,\,\,\,C}\right)\,,\n{A6.IV}\non\\\\
&&\hspace{-0.3cm}R_{kABC}=\kappa^{-1}({\cal S}_{AB;C}-{\cal S}_{AC;B})\,,\n{A7.IV}\\
&&\hspace{-0.3cm}R_{ABCD}={\cal R}_{ABCD}+{\cal S}_{AD}{\cal S}_{BC}-{\cal S}_{AC}{\cal S}_{BD}\,,\n{A8.IV}
\ea
where ${\cal R}_{ABCD}$ is the Riemann tensor of a $(d-2)$-dimensional surface $\Sigma_k$. Here and in what follows, the semicolon stands for the covariant derivative defined with respect to the $(d-2)$-dimensional metric $h_{AB}$, and ${\cal S}_{AB}$ is the  extrinsic curvature of a surface $\Sigma_k$ defined as
\be\n{A9.IV}
{\cal S}_{AB}\equiv\frac{1}{2}\kappa h_{AB,k}\,.
\ee
For the metric \eq{A4.IV} we derive
\ba
\hspace{-0.3cm}k_{|kk}&=&\kappa^{-1}\kappa_{,k}\hhh
k_{|Ak}=k_{|kA}=\kappa^{-1}\kappa_{,A}\,,\n{4.a}\\ 
\hspace{-0.3cm}k_{|AB}&=&\kappa{\cal S}_{AB}\hhh
k_{|a}^{~|a}=\kappa (\kappa_{,k}+{\cal S})\hhh{\cal S}\equiv {\cal S}_{A}^{\,\,\,\,A}\,.\n{4.d}
\ea
Using the Gauss and Codazzi relations \eq{A6.IV}-\eq{A8.IV} together with the relations \eq{4.a} and \eq{4.d}, we can write the components of the $d$-dimensional Ricci tensor \eq{R11} in terms of geometric quantities of a $(d-2)$-dimensional surface $\Sigma_k$ as follows:
\ba
&&\hspace{-0.25cm}R_{tt}=k\kappa(\kappa_{,k}+{\cal S})\,,\n{5.a}\\
&&\hspace{-0.25cm}R_{kk}=-\kappa^{-1}\left({\cal S}_{,k}+(\kappa^{-1})_{;A}^{~;A}+\kappa^{-1}{\cal S}_{A}^{~B}{\cal S}_{B}^{~A}+k^{-1}\kappa_{,k}\right)\,,\n{5.b}\non\\\\
&&\hspace{-0.25cm}R_{Ak}=-\kappa^{-1}\left({\cal S}_{,A}-{\cal S}_{A~;B}^{~B}+k^{-1}\kappa_{,A}\right)\,,\n{5.c}\\
&&\hspace{-0.25cm}R_{AB}={\cal R}_{AB}-{\cal S}{\cal S}_{AB}-\kappa(\kappa^{-1})_{;AB}-\kappa h_{AC}{\cal S}_{B~,k}^{~C}\non\\
&&\hspace{0.6cm}-\,k^{-1}\kappa\,{\cal S}_{AB}\,\n{5.d}.
\ea
Here ${\cal R}_{AB}$ is the Ricci tensor of a $(d-2)$-dimensional surface $\Sigma_k$.

Let us now define the electromagnetic field tensor $F_{\alpha\beta}$ in the static space-time \eq{A4.IV}. We consider the electrostatic $d$-potential 
\be
A_{\mu}=-\Phi \delta^t_{~\mu},
\ee 
where $\Phi=\Phi(k, x^A)$ is an electrostatic potential. The corresponding components of the electromagnetic field tensor read
\ba
\hspace{-0.3cm}F_{at}=-F_{ta}=-\Phi_{,a}\hhh F_{ab}=0\,.\n{1.4a}
\ea
They give
\ba
&&\hspace{-0.3cm}{F}^2=-2k^{-2} \Phi_{,a}\Phi^{,a}=-2k^{-2}(\kappa^2\Phi_{,k}^2+\Phi_{,A}\Phi^{,A})\,,\n{1.4b} \\
&&\hspace{-0.3cm}F_{t}^{~\alpha}F_{t\alpha}=\Phi_{,a}\Phi^{,a}=(\kappa^2\Phi_{,k}^2+\Phi_{,A}\Phi^{,A})\,,\n{1.4bba}\\
&&\hspace{-0.3cm}F_{a}^{~\alpha}F_{b\alpha}=-k^{-2}\Phi_{,a}\Phi_{,b}\,.\n{1.4c}
\ea
The dilaton field $\varphi$ does not depend on time, i.e. $\varphi=\varphi(k, x^A)$.

\begin{widetext} 
Using Eqs. \eq{5.a}-\eq{1.4c} we can present the EMdL equations \eq{1.1c}-\eq{1.3b} in the following form:
\newline the Maxwell equation \eq{1.1c} 
\ba
\hspace{-1cm}k(k^{-1}\sqrt{h}\kappa e^{a\varphi} \Phi_{,k})_{,k}+(\kappa^{-1}\sqrt{h} e^{a\varphi}\Phi^{,A})_{,A}=0\,, \n{1.5ce}\ea
the Klein-Gordon equation \eq{1.1d}
\ba
\hspace{-1cm}(k\kappa \sqrt{h}\varphi_{,k})_{,k}+(k\kappa^{-1}\sqrt{h}h^{AB}\varphi_{,A})_{,B}=-\frac{a}{2}k^{-1}\kappa^{-1}\sqrt{h}e^{a\varphi}(\kappa^2\Phi_{,k}^2+\Phi_{,C}\Phi^{,C})-2b\Lambda k\kappa^{-1}\sqrt{h}e^{-b\varphi}\,,\n{1.5cf}
\ea
and the Einstein equation \eq{1.3b}
\ba
&&\hspace{3cm}k\kappa(\kappa_{,k}+{\cal S})=\frac{d-3}{2(d-2)}\,e^{a\varphi}(\kappa^2\Phi_{,k}^2+\Phi_{,A}\Phi^{,A})-\frac{2\Lambda k^2}{d-2}e^{-b\varphi}\,,\n{1.5ca}\\
&&\hspace{-1cm}-\kappa^{-1}({\cal S}_{,k}+(\kappa^{-1})_{;A}^{~;A}+\kappa^{-1}{\cal S}_{A}^{~B}{\cal S}_{B}^{~A}+k^{-1}\kappa_{,k})=\frac{1}{2}\,\varphi_{,k}\varphi_{,k}-\frac{k^{-2}\kappa^{-2}}{2(d-2)}e^{a\varphi}\left[(d-3)\kappa^2\Phi_{,k}^2-\Phi_{,A}\Phi^{,A}\right]+\frac{2\Lambda\kappa^{-2}}{d-2}\,e^{-b\varphi},\n{1.5cb}\\
&&\hspace{3cm}-\kappa^{-1}({\cal S}_{,A}-{\cal S}_{A~;B}^{~B}+k^{-1}\kappa_{,A})=\frac{1}{2}\,\varphi_{,A}\varphi_{,k}-\frac{1}{2}k^{-2}e^{a\varphi}\Phi_{,k}\Phi_{,A}\,,\n{1.5cc}\\
&&\hspace{-1cm}{\cal R}_{AB}-{\cal S}{\cal S}_{AB}-\kappa(\kappa^{-1})_{;AB}-\kappa h_{AC}{\cal S}_{B,k}^{~C}-k^{-1}\kappa\,{\cal S}_{AB}=\frac{1}{2}\,\varphi_{,A}\varphi_{,B}-\frac{1}{2}k^{-2}e^{a\varphi}\Phi_{,A}\Phi_{,B}\non\\
&&\hspace{7cm}+\frac{k^{-2}}{2(d-2)}e^{a\varphi}(\kappa^2\Phi_{,k}^2+\Phi_{,C}\Phi^{,C})h_{AB}+\frac{2\Lambda}{d-2}e^{-b\varphi} h_{AB}\,.\n{1.5cd}
\ea
Here $h\equiv det(h_{AB})$.
Taking the trace of Eq. \eq{1.5cd} and using Eqs. \eq{1.5ca} and \eq{1.5cb} we derive
\ba
{\cal R}-{\cal S}^2+{\cal S}_A^{~B}{\cal S}_B^{~A}+2k^{-1}\kappa \kappa_{,k}&=&\frac{1}{2}(\varphi_{,A}\varphi^{,A}-\kappa^2\hnh\varphi_{,k}\varphi_{,k})+\frac{k^{-2}}{2(d-2)}e^{a\varphi}\left[(3d-8)\kappa^2\Phi_{,k}^2+(d-4)\Phi_{,A}\Phi^{,A}\right]\non\\
&+&\frac{2d-8}{d-2}\Lambda e^{-b\varphi}\,.\n{6}
\ea 
\end{widetext}
In Sec. V we construct approximate solutions to these equations in the vicinity of space-time Killing horizon ${\cal H}$.
\section{Space-time Curvature invariants}
In this section we present expressions of the space-time curvature invariants corresponding to the static space-time \eq{A4.IV} such as the Ricci scalar $R\equiv g^{\alpha\beta}R_{\alpha\beta}$, the square of the Ricci tensor $R_{\alpha\beta}R^{\alpha\beta}\,$, and the Kretschmann scalar ${\cal K}\equiv R_{\alpha\beta\gamma\delta}\hspace{0.09cm}R^{\alpha\beta\gamma\delta}$, in terms of geometric quantities of a $(d-2)$-dimensional surface $\Sigma_k$. 
Using Eqs. \eq{5.a}-\eq{5.d} we derive the $d$-dimensional Ricci scalar 
 \ba\n{Ri}
R&=&-2\left(k^{-1}\kappa\kappa_{,k}+k^{-1}\kappa{\cal S}+\kappa {\cal S}_{,k}+\kappa (\kappa^{-1})_{;A}^{~;A}\right)\non\\
&-&{\cal S}_{A}^{~B}{\cal S}_B^{~A}+{\cal R}-{\cal S}^2\,,
\ea
where ${\cal R}={\cal R}_A^{~A}$ is the Ricci scalar defined on $\Sigma_k$. 
The square of the $d$-dimensional Ricci tensor reads
\ba
&&\hspace{-0.3cm}R_{\alpha\beta}R^{\alpha\beta}=k^{-2}\kappa^2(\kappa_{,k}+{\cal S})^2\non\\
&&\hspace{-0.39cm}+\,\kappa^2\left({\cal S}_{,k}+(\kappa^{-1})_{;A}^{~;A}
+\kappa^{-1}{\cal S}_A^{~C}{\cal S}_C^{~A}+k^{-1}\kappa_{,k}\right)^2\non\\
&&\hspace{-0.39cm}+\,2\left({\cal S}_{,A}-{\cal S}_{A~;B}^{~B}+k^{-1}\kappa_{,A}\right)\left({\cal S}^{,A}-{\cal S}^{AC}_{~;C}+k^{-1}\kappa^{,A}\right)\non\\
&&\hspace{-0.39cm}+\left[{\cal R}_{AB}-{\cal S}{\cal S}_{AB}-\kappa(\kappa^{-1})_{;AB}-\kappa h_{AC}{\cal S}_{B~,k}^{~C}-k^{-1}\kappa {\cal S}_{AB}\right]\non\\
&&\hspace{-0.39cm}\times\left[{\cal R}^{AB}-{\cal S}{\cal S}^{AB}-\kappa(\kappa^{-1})^{;AB}-\kappa h^{BD}{\cal S}_{D~,k}^{~A}-k^{-1}\kappa {\cal S}^{AB}\right].\non\\\n{RT}
\ea
To calculate the Kretschmann scalar we need to derive the $d$-dimensional Riemann tensor components. Using Eqs. \eq{CC1a}-\eq{S4t} for the static space-time \eq{A1.IV} we derive
\ba\n{A16.IV}
R_{attb}=-kk_{|ab} \hhh R_{tabc}=0\hhh
R_{abcd}={\bar R}_{abcd}\,.
\ea
Then the Kretschmann scalar of the static space-time \eq{A4.IV} can be presented in the following form:
\ba\n{A17.IV}
\hspace{-0.5cm}{\cal K}&\equiv&R_{\alpha\beta\gamma\delta}\hspace{0.09cm}\hspace{-0.09cm}R^{\alpha\beta\gamma\delta}=4k^{-2}k_{|ab}k^{|ab}+4R_{kABC}R^{kABC}\non\\
&+&4R_{AkkB}R^{AkkB}+R_{ABCD}R^{ABCD}\,.
\ea
For $d\geqslant 5$ the $(d-2)$-dimensional Riemann tensor components ${\cal R}_{ABCD}$ corresponding to the metric $h_{AB}$ can be presented as follows (see, e.g., \cite{Ch}, p. 32):
\ba\n{A19.IV}
{\cal R}_{ABCD}&=&{\cal C}_{ABCD}+\frac{1}{(d-4)}(h_{AC}{\cal R}_{BD}\non\\
&+&h_{BD}{\cal R}_{AC}-h_{AD}{\cal R}_{BC}-h_{BC}{\cal R}_{AD})\non\\
&-&\frac{1}{(d-3)(d-4)}\hnh{\cal R}\left(h_{AC}h_{BD}-h_{AD}h_{BC}\right)\,,\non\\
\ea
where ${\cal C}_{ABCD}$ is the Weyl tensor defined on $\Sigma_k$. For $d=5$ the Weyl tensor ${\cal C}_{ABCD}$ vanishes identically. This expression implies
\ba\n{A20.IV}
{\cal R}_{ABCD}{\cal R}^{ABCD}&=&{\cal C}_{ABCD}{\cal C}^{ABCD}+\frac{4}{d-4}{\cal R}_{AB}{\cal R}^{AB}\non\\
&-&\frac{2}{(d-3)(d-4)}\hnh{\cal R}^2\,.
\ea
For  $d=4$ the 2-dimensional Reimann tensor components corresponding to the metric $h_{AB}$ have the following form:
\be
{\cal R}_{ABCD}=\frac{1}{2}(h_{AC}h_{BD}-h_{AD}h_{BC}){\cal R}\,.\n{Rin4}
\ee
This expression implies 
\ba
{\cal R}_{ABCD}{\cal R}^{ABCD}={\cal R}^2\hhh{\cal R}_{AB}=\frac{1}{2}h_{AB}{\cal R}\,.\n{Eq4d}
\ea
Using the expressions \eq{A17.IV}-\eq{Eq4d} we derive the Kretschmann scalar for the static $d$-dimensional ($d\geqslant 5$) space-time \eq{A4.IV}
\ba\n{A21.IV}
{\cal K}&=&4k^{-2}\kappa^2\left(\kappa_{,k}^2+2\kappa^{-2}\kappa_{,A}\kappa^{,A}+{\cal S}_{AB}{\cal S}^{AB}\right)\non\\
&+&4\kappa^2\left[h_{AC}h^{BD}{\cal S}_{B~,k}^{~C}{\cal S}_{D~,k}^{~A}
+2\kappa^{-1}{\cal S}_{AC}{\cal S}^{AB}{\cal S}_{B~,k}^{~C}\right.\non\\
&+&2\kappa^{-1}(\kappa^{-1})_{;AB}{\cal S}^{A}_{~C}{\cal S}^{BC}
+2(\kappa^{-1})_{;A}^{~;B}{\cal S}_{B~,k}^{~A}\non\\
&+&\left(\kappa^{-1})_{;AB}(\kappa^{-1})^{;AB}\right]
+8{\cal S}^{AB;C}({\cal S}_{AB;C}-{\cal S}_{AC;B})\non\\
&+&{\cal C}_{ABCD}{\cal C}^{ABCD}+2{\cal C}_{ABCD}{\cal S}^{AD}{\cal S}^{BC}\non\\
&+&2{\cal C}_{ABCD}{\cal S}^{AC}{\cal S}^{BD}+\frac{4}{d-4}\left({\cal R}_{AB}{\cal R}^{AB}\right.\non\\
&-&\left.2{\cal S}{\cal S}^{AB}{\cal R}_{AB}+2{\cal R}_{AC}{\cal S}_B^{~C}{\cal S}^{AB}\right)\non\\
&-&\frac{2}{(d-3)(d-4)}{\cal R}\left({\cal R}-2{\cal S}^2+2{\cal S}_{AB}{\cal S}^{AB}\right)\non\\
&+&2{\cal S}_{AB}{\cal S}^{AB}{\cal S}_{CD}{\cal S}^{CD}+2{\cal S}_{AC}{\cal S}^{BC}{\cal S}_{BD}{\cal S}^{AD}\,,
\ea
and for the static $4$-dimensional space-time \eq{A4.IV} 
\ba 
{\cal K}&=&4k^{-2}\kappa^2\left(\kappa_{,k}^2+2\kappa^{-2}\kappa_{,A}\kappa^{,A}+{\cal S}_{AB}{\cal S}^{AB}\right)\non\\
&+&4\kappa^2\left[h_{AC}h^{BD}{\cal S}_{B~,k}^{~C}{\cal S}_{D~,k}^{~A}\right.
+2\kappa^{-1}{\cal S}_{AC}{\cal S}^{AB}{\cal S}_{B~,k}^{~C}\non\\
&+&2\kappa^{-1}(\kappa^{-1})_{;AB}{\cal S}^{A}_{~C}{\cal S}^{BC}
+2(\kappa^{-1})_{;A}^{~;B}{\cal S}_{B~,k}^{~A}\non\\
&+&\left.(\kappa^{-1})_{;AB}(\kappa^{-1})^{;AB}\right]
+8{\cal S}^{AB;C}({\cal S}_{AB;C}-{\cal S}_{AC;B})\non\\
&+&{\cal R}^2+2{\cal R}\left({\cal S}_{AB}{\cal S}^{AB}-{\cal S}^2\right)+2{\cal S}_{AB}{\cal S}^{AB}{\cal S}_{CD}{\cal S}^{CD}\non\\
&+&2{\cal S}_{AC}{\cal S}^{BC}{\cal S}_{BD}{\cal S}^{AD}\,.\n{A21.IVb}
\ea
The expressions \eq{Ri}, \eq{RT} and \eq{A21.IV}, \eq{A21.IVb} define the space-time curvature invariants everywhere in the space-time \eq{A4.IV}. In the next section we calculate the space-time curvature invariants on its Killing horizon ${\cal H}$.
\section{Geometry Near the Horizon}
In this section we derive asymptotic behavior of the metric \eq{A4.IV}, its related geometric quantities, the electrostatic potential $\Phi$, and the dilaton field $\varphi$ near the Killing horizon ${\cal H}$ ($k=0$).
Equations \eq{1.4b}, and \eq{1.5ce}-\eq{6}, together with the definition \eq{A9.IV} provide a complete system of equations for determining such an asymptotic behavior. Assuming that the metric functions $\kappa$ and $h_{AB}$, and the fields $\Phi$ and $\varphi$ are real analytic functions of $k$ in the vicinity of ${\cal H}$, we construct the following series expansions:
\ba\n{A22.IV}
&&\hspace{1.5cm}{\cal A}=\sum_{n\geqslant0}{\cal A}^{[n]}k^{n}\,,\non\\
&&{\cal A}=\{\kappa, h_{AB},{\cal S}_{AB},{\cal S},{\cal R}_{AB},{\cal R},\Phi,F^2,\varphi\}\,,
\ea
which converge in the vicinity of ${\cal H}$. Here the first term ${\cal A}^{[0]}$ corresponds to the value of ${\cal A}$ calculated on the horizon, i.e. ${\cal A}^{[0]}\equiv{\cal A}\rvert_{\cal H}$. We have two types of quantities, $d$-dimensional and $(d-2)$-dimensional which are defined on $\Sigma_k$ surfaces. If ${\cal A}$ is a $d$-dimensional quantity, e.g. $F^2$ or ${\cal K}$, then ${\cal A}\rvert_{\cal H}$ is its value on the $(d-1)$-dimensional horizon ${\cal H}$ defined by $k=0$; if ${\cal A}$ is a $(d-2)$-dimensional quantity, e.g. ${\cal S}_{AB}$ or ${\cal R}_{AB}$, then ${\cal A}\rvert_{\cal H}$ is its value on the $(d-2)$-dimensional horizon surface $\Sigma_k$ defined by $t=const$, $k=0$.

A substitution of the expansions \eq{A22.IV} into the EMdL equations \eq{1.5ce}-\eq{6} gives equations for the expansion coefficients ${\cal A}^{[n]}$. The lowest order terms corresponding to $k^{-2}$ and $k^{-1}$ give
\ba
&&\kappa^{[0]}\Phi^{[1]}\sqrt{h}^{[0]}e^{a\varphi^{[0]}}=0\,,\n{A}\\
&&\kappa^{[0]}=const\hhh \Phi^{[0]}=const\,,\n{B}\\
&&\kappa^{[0]}\kappa^{[1]}=0\hhh \kappa^{[0]}{\cal S}_{AB}^{[0]}=0\,.\n{C}
\ea
Equation \eq{B} means that the horizon surface gravity and the electrostatic potential are constant on the horizon. Assuming that $\kappa^{[0]}\neq0$ and $\sqrt{h}^{[0]}\neq0$, equations \eq{A} and \eq{C} imply
\be
\Phi^{[1]}=0\hhh \kappa^{[1]}=0\hhh {\cal S}_{AB}^{[0]}=0\,.\n{D}
\ee
According to the definition \eq{A9.IV}, we have 
\ba
\hspace{-0.6cm}{\cal S}_{AB}&=&{\cal S}_{AB}^{[0]}+{\cal S}_{AB}^{[1]}k+...\nonumber\\
&=&\frac{1}{2}\kappa^{[0]}h_{AB}^{[1]}+\frac{1}{2}(\kappa^{[1]}h_{AB}^{[1]}+2\kappa^{[0]}h_{AB}^{[2]})k+...~.\n{E}
\ea
Thus, equations \eq{D} and \eq{E} imply
\be
h_{AB}^{[1]}=0\,.\n{F}
\ee
According to Eq. \eq{E} for $\kappa^{[0]}=0$ we again have ${\cal S}_{AB}^{[0]}=0$. Thus, in both cases $\kappa^{[0]}=const\neq 0$ and $\kappa^{[0]}=0$ the horizon surface is a totally geodesic surface.

The case $\kappa^{[0]}=const\neq0$ corresponds to a non-degenerate Killing horizon, while the case $\kappa^{[0]}=0$ corresponds to a degenerate (extremal) Killing horizon. In what follows, we shall consider these cases separately.
\subsection{Non-degenerate Horizon}
The higher order expansion coefficients ${\cal A}^{[n]}$ corresponding to $\kappa^{[0]}=const\neq0$ can be derived by plugging the expansions \eq{A22.IV} into the EMdL equations \eq{1.5ce}-\eq{6} and by using Eqs. \eq{B}, \eq{D}-\eq{F}, together with Eq. \eq{1.4b}.  We derive the following first and second order expansion coefficients:
\begin{widetext}
\ba
&&\kappa^{[2]}=(4\kappa^{[0]})^{-1}\left(\frac{1}{2}\varphi_{~,A}^{[0]}\varphi^{[0],A}-{\cal R}^{[0]}-\frac{3d-8}{4(d-2)} e^{a\varphi^{[0]}}{F}^{2[0]}\right.
\left.+\frac{2d-8}{d-2}\Lambda e^{-b \varphi^{[0]}}\right) \,,\n{8.c}\\
&&h_{AB}^{[2]}=-(\sqrt{2}\kappa^{[0]})^{-2}\left(\frac{1}{2}\varphi_{~,A}^{[0]}\varphi_{~,B}^{[0]}-{\cal R}_{AB}^{[0]}-\frac{e^{a\varphi^{[0]}}{F}^{2[0]}}{4(d-2)}h_{AB}^{[0]}+\frac{2\Lambda e^{-b\varphi^{[0]}}}{d-2}h_{AB}^{[0]}\right)\,,\n{8.db}\\
&&{\cal S}_{AB}^{[2]}=0
\hhh~~~\Phi^{[2]}=\pm\frac{\sqrt{-{F}^{2[0]}}}{2\sqrt{2}\kappa^{[0]}}\,,\n{8.a}\\
&&\varphi^{[1]}=0\hhh\varphi^{[2]}=-(\kappa^{[0]})^{-2}\left(\frac{1}{4}\varphi^{[0]~;A}_{~~;A}-\frac{a}{16}e^{a\varphi^{[0]}}{F}^{2[0]}+b\Lambda e^{-b\varphi^{[0]}}\right).\n{8.df}
\ea
\end{widetext}
As we said in the Introduction, we shall consider only secondary scalar hair induced by the electrostatic field and Liouville potential. According to the expression \eq{8.df}, we have to put 
\be
\varphi^{[0]}=const\,. \n{h62}
\ee 
Hence, all the terms in the expressions above which involve derivatives of $\varphi^{[0]}$ vanish. The condition \eq{h62} is necessary condition which excludes non-zero scalar kinetic term in the action \eq{Action2} for vanishing electrostatic field and the Liouville potential. 

The derived expansion coefficients define the metric and the fields near non-degenerate horizon of a general static $d$-dimensional analytic solution to the EMdL equations. Given the coupling constants $a$, $b$, and $\Lambda$, such a solution is defined in terms of the three constants $\kappa^{[0]}$, $\varphi^{[0]}$, and $\Phi^{[0]}$, and $1+(d-1)(d-2)/2$ functions ${F}^{2[0]}$ and $h_{AB}^{[0]}$. 

Let us now calculate the space-time curvature invariants \eq{Ri}, \eq{RT}, and \eq{A21.IV}, \eq{A21.IVb} on a non-degenerate horizon using the expansion coefficients derived above. 
Substituting the expansions \eq{A22.IV} into the expressions \eq{Ri} and \eq{RT} and taking the terms corresponding to $k^0$ we derive the Ricci scalar and the square of the Ricci tensor calculated on the horizon ${\cal H}$ 
\ba
&&\hspace{-1cm}{R}|_{\cal H}=\frac{d-4}{4(d-2)}e^{a\varphi^{[0]}}F^{2[0]}+\frac{2d\Lambda}{d-2} e^{-b\varphi^{[0]}},\n{Riex}\\
&&\hspace{-1cm}{ R}_{\alpha\beta}\hspace{0.01cm}{R}^{\alpha\beta}|_{\cal H}=
\frac{(2d^2-11d+16)}{16(d-2)^2}e^{2a\varphi^{[0]}}(F^{2[0]})^2\non\\
&&\hspace{-0.5cm}+\frac{(d-4)\Lambda}{(d-2)^2}e^{(a-b)\varphi^{[0]}}F^{2[0]}+\frac{4d\Lambda^2}{(d-2)^2}e^{-2b\varphi^{[0]}}.\n{RTex}
\ea
Note that ${R}|_{\cal H}$ is negative for $\Lambda<0$.

To calculate the Kretschmann scalar ${\cal K}$ on the horizon we substitute the expansions \eq{A22.IV} into the expressions \eq{A21.IV} and \eq{A21.IVb}, and take the terms corresponding to $k^0$. For a $d$-dimensional $(d\geqslant 5$) static space-time we derive
\ba\n{A23.IV}
{\cal K}\rvert_{{\cal H}}&=&{\cal C}_{ABCD}^{[0]}{\cal C}^{ABCD[0]}+\frac{2(d-2)}{(d-4)}{\cal R}^{[0]}_{AB}{\cal R}^{AB[0]}\non\\
&+&\frac{(d-2)(d-5)}{(d-4)(d-3)}({\cal R}^{[0]})^2\non\\
&+&\left(\frac{3}{2} F^{2[0]}e^{a\varphi^{[0]}}-4\Lambda e^{-b\varphi^{[0]}}\right){\cal R}^{[0]}\non\\
&+&\frac{(9d^2-46d+60)}{16(d-2)^2}e^{2a\varphi^{[0]}}(F^{2[0]})^2\non\\
&-&\frac{(3d^2-18+28)}{(d-2)^2}\Lambda  e^{(a-b)\varphi^{[0]}}F^{2[0]}\non\\
&+&\frac{4(d^2-6d+12)}{(d-2)^2}\Lambda^2 e^{-2b\varphi^{[0]}},
\ea 
where  ${\cal C}_{ABCD}^{[0]}$ is the Weyl tensor corresponding to the horizon surface. For $d=5$ we have ${\cal C}_{ABCD}=0$. The Kretschmann scalar calculated on the horizon of a 4-dimensional static space-time is 
\ba
{\cal K}\rvert_{{\cal H}}&=&3({\cal R}^{[0]})^2+\left(\frac{3}{2}F^{2[0]}e^{a\varphi^{[0]}}-4\Lambda e^{-b\varphi^{[0]}}\right){\cal R}^{[0]}\non\\
&+&\frac{5}{16}e^{2a\varphi^{[0]}}(F^{2[0]})^2-F^{2[0]}\Lambda e^{(a-b)\varphi^{[0]}}\non\\
&+&4\Lambda^2 e^{-2b\varphi^{[0]}}.\n{A23.IVb}
\ea
The expressions \eq{A23.IV} and \eq{A23.IVb} are generalizations of the known expressions \eq{A24.IV} and \eq{A24.IVb}. According to the expressions \eq{Riex}-\eq{A23.IVb}, if the fields $F^2$ and $\varphi$ are finite on a non-degenrate horizon and if the horizon surface is regular, i.e. ${\cal R}^{[0]}$, ${\cal R}^{[0]}_{AB}{\cal R}^{AB[0]}$, and ${\cal C}_{ABCD}^{[0]}{\cal C}^{ABCD[0]}$ are finite, then the horizon is regular.  

Let us mention a few examples of static solutions of the EMdL model which have non-degenerate Killing horizon. Exact black hole solutions to the EMdL equations with non-degenerate horizon can be found in \cite{7} and \cite{Ortin} (see p. 350). Spherically symmetric solutions representing black holes and black strings without a Liouville potential were derived in \cite{Jutta1, Jutta2}. Distorted, axisymmetric, charged, 4-dimensional dilaton black holes were constructed in \cite{10}. The $d$-dimensional Reissner-Nordstr\"om solution with a cosmological constant was derived in \cite{13}. Distorted, static, axisymmetric, four and $5$-dimensional vacuum and $4$-dimensional electrovacuum black holes were studied in e.g. \cite{FS1, Geroch, ASP} and \cite{Bre2, Step, AFS}, respectively. 
Note that the $d$-dimensional Rindler space-time corresponds to $\kappa=const$, $\varphi=\Phi=\Lambda=0$, and the flat Riemannian metric $h_{AB}$.
\subsection{Extremal Horizon}
Let us now consider the extremal case $\kappa^{[0]}=0$. Plugging the expansions \eq{A22.IV} into the EMdL equations \eq{1.5ce}-\eq{6} we derive 
\ba
\kappa^{[1]}=const\hhh \Phi^{[1]}=const\,.\n{kconst}
\ea
The constant values of $\kappa^{[1]}$ and $\Phi^{[1]}$ are calculated in terms of the coupling constants $a$, $b$, and $\Lambda$, the electrostatic field potential $\Phi^{[0]}$, and the dilaton field $\varphi^{[0]}$ from the following equations:
\ba
&&\hspace{-0.33cm}4\Lambda e^{-b\varphi^{[0]}}+(\kappa^{[1]})^2\left(2(d-2)-(d-3)(\Phi^{[1]})^2e^{a\varphi^{[0]}}\right)=0,\non\\\n{68a}\\
&&\hspace{-0.33cm}2\varphi^{[0]~;A}_{~~;A}+a(\kappa^{[1]})^2(\Phi^{[1]})^2e^{a\varphi^{[0]}}+4b\Lambda e^{-b\varphi^{[0]}}=0\,.\n{68b}
\ea
The first equation implies that for general values of the coupling constants and $\kappa^{[1]}$, $\Phi^{[1]}$ we have 
\be
\varphi^{[0]}=const\,. \n{h70}
\ee
Thus, in Eq. \eq{68b} the first term vanishes. Note that in the extremal case the condition \eq{h70} follows directly from the EMdL equations, while in the case of a non-degenerate horizon the same condition \eq{h62} is imposed by hand, in accordance with the secondary hair condition. 

The Ricci tensor of the horizon surface is defined by the following expression:
\ba
{\cal R}_{AB}^{[0]}&=&\frac{1}{2(d-2)}\left(4\Lambda e^{-b\phi^{[0]}}+(\kappa^{[1]})^2(\Phi^{[1]})^2e^{a\varphi^{[0]}}\right)h_{AB}^{[0]}.\non\\
\n{68c}
\ea
According to this expression, the extremal horizon surface is an Einstein space. It was shown in \cite{Tod} that spatially compact extremal horizons in $d$-dimensional ($d\geqslant 4$) adS space-times are compact Einstein spaces of negative curvature. 

The electromagnetic field invariant \eq{1.4b} calculated on the extremal horizon is given by
\be
F^{2[0]}=-2(\kappa^{[1]})^2(\Phi^{[1]})^2.\n{68d}
\ee
According to the expressions \eq{68c} and \eq{68d}, for 
\be
\Lambda<\frac{1}{8}F^{2[0]} e^{(a+b)\phi^{[0]}}\,
\ee
the extremal horizon surface has negative curvature.

Solving Eqs. \eq{68a} and \eq{68b} for $\kappa^{[1]}$ and $\Phi^{[1]}$, we can derive the Ricci tensor \eq{68c}, the electromagnetic field invariant \eq{68d}, and the space-time curvature invariants $R$, $R_{\alpha\beta}R^{\alpha\beta}$, and ${\cal K}$ calculated on the extremal horizon. 

According to Eqs. \eq{68a}-\eq{68d}, the metric and the fields depend on values of the coupling constants $a$, $b$, and $\Lambda$. There are six cases, which we shall consider separately. 
\subsubsection{Case $a\neq0$, $b\neq0$, $\Lambda\neq0$ }
In this case we have
\ba\n{74AB}
&&\hspace{-0.25cm}(\kappa^{[1]})^2=-2\Lambda e^{-b\varphi^{[0]}}\frac{[a+b(d-3)]}{a(d-2)}\,,\non\\
&&\hspace{-0.25cm} (\Phi^{[1]})^2=\frac{2b(d-2)e^{-a\varphi^{[0]}}}{a+b(d-3)}\hhh F^{2[0]}=\frac{8b\Lambda}{a}e^{-(a+b)\varphi^{[0]}},\non\\
&&\hspace{-0.25cm}{\cal R}_{AB}^{[0]}=\frac{2\Lambda (a-b)}{a(d-2)}e^{-b\varphi^{[0]}}h_{AB}^{[0]},~
{\cal R}^{[0]}=\frac{2\Lambda }{a}(a-b)e^{-b\varphi^{[0]}}.\non\\
\ea
A solution with real values of $\kappa^{[1]}$ and $\Phi^{[1]}$ exists for $\Lambda>0$, when $a>0$ and $b<-a/(d-3)$, or $a<0$ and $b>-a/(d-3)$, and for $\Lambda<0$, when $a>0$ and $b>0$, or $a<0$ and $b<0$.
In the case $a=b\neq0$ the extremal horizon surface is a Ricci flat Riemannian surface. Some examples of the {\em Case 1} are discussed in \cite{16}, and in the case of 4-dimensional space-times in \cite{18} and \cite{17}.

Substituting the expansions \eq{A22.IV} into the expressions \eq{Ri}, \eq{RT}, and \eq{A21.IV}, \eq{A21.IVb} and taking the terms corresponding to $k^0$ we derive the following space-time curvature invariants calculated on the extremal horizon: the Ricci scalar 
\ba
&&R|_{{\cal H}}=\frac{2\Lambda e^{-b\varphi^{[0]}}}{a(d-2)}[ad+b(d-4)]\,,
\ea
the square of the Ricci tensor 
\ba
&&\hspace{-1.2cm}R_{\alpha\beta}R^{\alpha\beta}|_{{\cal H}}=\frac{4\Lambda^2 e^{-2b\varphi^{[0]}}}{a^2(d-2)^2}\non\\
&&\hspace{-0.5cm}\times[a^2d+2ab(d-4)+b^2(2d^2-11d+16)]\,,
\ea
the Kretschmann scalar ($d\geqslant 5$)
\ba
{\cal K}|_{\cal H}&=&\frac{8\Lambda^2e^{-2b\varphi^{[0]}}}{a^2(d-2)}\left(\frac{2[a+b(d-3)]^2}{d-2}+\frac{(a-b)^2}{d-3}\right)\non\\
&+&{\cal C}_{ABCD}^{[0]}{\cal C}^{ABCD[0]}\,,
\ea
where for $d=5$ we have ${\cal C}_{ABCD}=0$, and the Kretschmann scalar ($d=4$)
\ba
{\cal K}|_{\cal H}&=&\frac{8\Lambda^2(a^2+b^2)}{a^2}e^{-2b\varphi^{[0]}}.
\ea
\subsubsection{Case $a\neq 0$, $b=0$, $\Lambda\neq 0$}
This case follows from the previous case by setting $b=0$. According to the expression \eq{74AB} for $\kappa^{[1]}$,
a real solution exists only for $\Lambda<0$. The electromagnetic field invariant vanishes on the horizon $F^{2[0]}=0$.  Note also that in this case the space-time curvature invariants: the Ricci scalar, the square of the Ricci tensor, and the Kretschmann scalar calculated on the extremal horizon do not depend on the dilaton coupling constant $a$. 

\subsubsection{Case $a\neq 0$, $\Lambda=0$}
According to Eqs. \eq{68a} and \eq{68b}, in this case we have $\kappa^{[1]}=0$. Note that in contrast to the general {\em Case 1}, in this particular case we have to set by hand $\varphi^{[0]}=const$ in Eq. \eq{68b}, in accordance with the secondary scalar hair condition. Calculating the higher order terms in the expansions \eq{A22.IV} we derive
\be
\kappa^{[2]}=\kappa^{[3]}=...=0\hhh \Phi^{[2]}=const\hhh \Phi^{[3]}=const,...~.
\ee
Here the constant values of $\Phi^{[n]}$, $n\geqslant 1$ are arbitrary and not related to each other. Thus, the electrostatic potential $\Phi$ is not defined and $\kappa\equiv0$. This implies that there are no extremal horizons corresponding to this case within the analytic expansions \eq{A22.IV}. Known static and spherically symmetric solutions corresponding to this case have singular extremal horizon, see e.g. \cite{7} and \cite{Ortin} (p. 350).
\subsubsection{Case $a=b=0$, $\Lambda\neq0$}
In this case the secondary scalar hair condition implies that Eq. \eq{68b} is an identity. Using Eqs. \eq{68a} and \eq{68d} we can express $\kappa^{[1]}$ and $\Phi^{[1]}$ in terms of $F^{2[0]}=const$ as follows:
\ba
&&(\kappa^{[1]})^2=-\frac{8\Lambda+(d-3)F^{2[0]}}{4(d-2)}\,,\\
&&(\Phi^{[1]})^2=\frac{2(d-2)F^{2[0]}}{8\Lambda+(d-3)F^{2[0]}}\,.
\ea
Note that a solution with real values of $\kappa^{[1]}$ and $\Phi^{[1]}$ exits only for $8\Lambda+(d-3)F^{2[0]}>0$.
Substituting these expressions into Eq. \eq{68c} we derive 
\ba
\hspace{-0.3cm}{\cal R}_{AB}^{[0]}=\frac{8\Lambda-F^{2[0]}}{4(d-2)}h_{AB}^{[0]}\hhh
{\cal R}^{[0]}=2\Lambda-\frac{1}{4}F^{2[0]}\,.
\ea
Note that electrically charged or neutral (a)dS black holes belong to this case (see, e.g. \cite{13}).

The space-time curvature invariants calculated on the extremal horizon  corresponding to this case are the following: 
the Ricci scalar \eq{Ri} 
\ba
&&R|_{{\cal H}}=\frac{8\Lambda d+(d-4)F^{2[0]}}{4(d-2)}\,,
\ea
the square of the Ricci tensor \eq{RT} 
\ba
R_{\alpha\beta}R^{\alpha\beta}|_{{\cal H}}&=&\frac{1}{16(d-2)^2}\biggl[64d\Lambda^2+16\Lambda F^{2[0]}(d-4)\non\\
&+&(F^{2[0]})^2(2d^2-11d+16)\biggl]\,,
\ea
the Kretschmann scalar \eq{A21.IV} ($d\geqslant 5$)
\ba
{\cal K}|_{{\cal H}}&=&{\cal C}_{ABCD}^{[0]}{\cal C}^{ABCD[0]}+\frac{1}{8(d-2)^2(d-3)}\non\\
&\times&\biggl[64\Lambda^2(3d-8)+16\Lambda F^{2[0]}(2d-5)(d-4)\non\\
&+&(F^{2[0]})^2(2d^3-18d^2+55d-56)\biggl]\,,
\ea
where for $d=5$ we have ${\cal C}_{ABCD}=0$, and
the Kretschmann scalar \eq{A21.IVb} ($d=4$)
\be
{\cal K}|_{\cal H}=8\Lambda^2+\frac{1}{8}(F^{2[0]})^2\,.
\ee
\subsubsection{Case $a=0$, $\Lambda=0$}
This case follows from the previous case by setting $\Lambda=0$. The $d$-dimensional extremal Reissner-Nordstr\"om black hole belongs to this case (see, e.g. \cite{13}). 

\subsubsection{Case $a=0$, $b\neq0$, $\Lambda\neq0$}
From Eqs. \eq{68b} and \eq{h70} we conclude that in this case no extremal horizons corresponding to the analytic expansions \eq{A22.IV} exist. 

The conditions presented in the cases above are necessary conditions for existence of static extremal horizons within the EMdL model. One should keep in mind that the external to such an extremal horizon space-time region may have singularities located beyond the radius of the convergence of the expansions \eq{A22.IV}. In addition, one has to rule out the points in the external region where $\kappa\equiv 0$, i.e. the points where gradient to  equipotential surfaces $\Sigma_k$ vanishes (see Eq. \eq{A5.IV}). At such points Israel's description becomes inapplicable. If at such a point space-time is regular, one may consider another coordinate system in the vicinity of this point. Moreover, the condition \eq{h70} may not imply the secondary hair condition as such, and one may need to impose an additional condition to ensure that the expansion coefficients $\varphi^{[n]}$, $n\geqslant 1$ vanish identically for vanishing electrostatic field and the Liouville potential. All these important issues can be addressed to non-degenerate horizons as well. Such issues are related to detailed and involved analysis of space-times at hand which we shall not undertake here. The interested reader can find explicit examples of near-horizon geometries of extremal static and stationary horizons in, e.g. \cite{Ha1,Ha1a,Ha2,Ha3}.

\section{Summary}
Let us summarize our results. We studied the geometric properties of static non-degenerate and degenerate (extremal) Killing horizons of arbitrary geometry and topology within the EMdL model under assumption that the metric functions and fields are analytic in the vicinity of the horizons. Such horizons have zero extrinsic curvature,  constant surface gravity, and the electrostatic potential is constant on such horizons. The presence of a dilaton field corresponding to the secondary scalar hair imposes an additional condition: the dilaton field should be constant on non-degenerate horizons. However, in the case of an extremal horizon this condition follows directly form the EMdL equations, except for the {\em Case 3}, which corresponds to zero Liouville potential, and for the {\em Case 4}, which corresponds to $a=b=0$. We derived the relations between space-time curvature invariants (the Ricci scalar, the square of the Ricci tensor, and the Kretschmann scalar) calculated on static EMdL horizons and the geometric quantities corresponding to the horizons surface. These relations are generalizations of the analogous known relations for horizons of static four and 5-dimensional vacuum and 4-dimensional electrovacuum space-times (see Eqs. \eq{A24.IV} and \eq{A24.IVb}). We have shown that all static extremal horizon surfaces of the EMdL model which correspond to the analytic expansions \eq{A22.IV} of the metric functions and fields are Einstein spaces, and presented the necessary conditions for existence of static extremal horizons within the EMdL model. In the case $a=b\neq0$ and $\Lambda\neq0$ surface of such an extremal horizon is Ricci flat. In contrast to a non-extremal horizon, in the general {\em Case 1} the electromagnetic field invariant $F^2$ calculated on the horizon is not an arbitrary constant, and is defined in terms of the coupling constants $a$, $b$, and $\Lambda$, and the dilaton field $\varphi$ calculated on the horizon. In the particular case $b=0$ ({\em Case 2}) corresponding to a negative cosmological constant $\Lambda<0$, the electromagnetic field invariant $F^2$ vanishes on the extremal horizon. It is interesting that in this case the space-time curvature invariants, i.e. the Ricci scalar, the square of the Ricci tensor, and the Kretschmann scalar, calculated on the extremal horizon do not depend on the dilaton coupling constant $a$. We also found that in the case of zero cosmological constant ({\em Case 3}) or vanishing dilaton coupling constant ({\em Case 6}) extremal horizons do not exist. Exact analytic solutions to the EMdL model support these necessary conditions. We believe that the approach presented here can be used for studying other models which contain more fields. This approach allows one to construct geometry near horizon of a static space-time at hand. Once such geometry is known, one can define the corresponding global space-time structure and analyze its properties. The geometric properties of the horizons presented here may be important for applications to holographic models as well as for understanding of properties of space-time horizons in general. 
\begin{acknowledgments}
The authors wish to thank Hari K. Kunduri for useful suggestions and for bringing our attention to the paper \cite{Tod}. One of the authors (A. A. S.) is grateful to the Natural Sciences and Engineering Research Council of Canada for the financial support.
\end{acknowledgments}

\end{document}